\documentclass[aps,prb,twocolumn,showpacs,preprintnumbers,amsmath,amssymb]{revtex4}
\usepackage{graphicx}
\usepackage{txfonts}

\begin{document}


\title{Small-angle interband scattering as the origin of the $T^{3/2}$ resistivity in MnSi}
\author{M. F. Smith}
 \email{msmith@nsrc.or.th}
\affiliation{%
National Synchrotron Research Center, Nakhon Ratchasima, Thailand,
30000 }
\date{\today}

\begin{abstract}
A possible explanation is given for the anomalous $T^{3/2}$
temperature dependence of the electrical resistivity of MnSi,
which is observed in the high-pressure paramagnetic state. The
unusual Fermi surface of MnSi includes large
open sheets that intersect along the faces of the cubic Brillouin
zone.  Close to these intersections, long-wavelength interband
magnetic spin fluctuations can scatter electrons from one sheet to
the other. The current relaxation rate due to such interband
scattering events is not reduced by vertex corrections as is that
for scattering from intraband ferromagnetic fluctuations.
Consequently, current relaxation proceeds in a manner similar to
that occurring in nearly antiferromagnetic metals, in which
low-temperature $T^{3/2}$ behavior is well known.  It is argued
that this type of non-Fermi-liquid behavior can, for a metal with
ferromagnetic fluctuations near Fermi sheet intersections, persist
over a much wider temperature range than it does in nearly
antiferromagnetic metals.
\end{abstract}

\pacs{72.15-v, 75.10.Lp, 75.40.Gb}

\maketitle

\section{Introduction}

Much recent attention has been paid to the itinerant helimagnet
MnSi\cite{juli97,doir03,pfei04,fisc04,jeon04,beli05,tewa05,kirk05,belib05,binz06,beli06}.
The ordered state occurs at low pressure, and is characterized by
a helical spiral with a period of 180 $\r{A}$, making the system
nearly ferromagnetic on atomic length scales. Current interest is
mainly in properties of the system observed at high pressure,
outside the helimagnetic dome in the $P-T$ phase diagram. For
example, at pressures above the critical value $P_C=14.6$ Kbar
required to kill the helimagnetism at $T=0$, the electrical
resistivity has been found\cite{doir03} to vary with temperature
as $\rho(T)=\rho_0+aT^{3/2}$. This non-Fermi liquid (NFL)
behavior, which is the subject of this article, persists down to
the lowest measured temperature of below 40 mK and up to the highest
applied pressure of greater than $2P_C$.  It is in apparent
disagreement with the standard picture of nearly magnetic
metals\cite{hert76,mill93,sach99} according to which the predicted
low-$T$ NFL resistivity \cite{mori85,lonz85} varies as $T^{5/3}$
close to a transition to ferromagnetism.  Also at pressures well above
$P_C$, neutron scattering measurements have revealed a novel type
of long-range magnetic order\cite{pfei04}.  A recent theoretical
proposal\cite{binz06} describes this so-called partial order
as a complex 3D spin configuration within a 
large (180$\r{A}$) cubic
supercell that is periodically repeated.  The variation of the direction of the 
local magnetization occurs on the length scale of the supercell 
so spins are aligned over distances significantly larger than the atomic spacing. 

The NFL behavior is seen both when the partial order is
present and when it is absent.  This suggests that the $T^{3/2}$ dependence might be explained
using a simple model that contains long-wavelength 
spin fluctuations, which appear to remain well above $P_C$, 
without worrying about the actual 
spin configuration.  Since there are no other known ferromagnets (or long wavelength
helimagnets) that exhibit $T^{3/2}$-resistivity in the
non-magnetic state\cite{doir03}, peculiarities of the MnSi band
structure also deserve attention. Recent theoretical band-structure
work\cite{jeon04} shows that the Fermi surface of MnSi has unusual
features related to large intersecting Fermi surface sheets that
provide ample phase space for small wavevector interband
scattering.

In this article I show that the observed $T^{3/2}$ behavior of the
electrical resistivity is expected if long-wavelength magnetic
fluctuations that couple spins on the intersecting Fermi surface
sheets play a significant role as scatterers.  This result follows
from the simplest model of a nearly-ferromagnetic system along
with consideration of key features in the MnSi Fermi surface.  The
main assumption is that MnSi has a tendency towards
long-wavelength interband magnetic correlations at pressures above
$P_C$.  I should emphasize that no attempt is made here to explain the magnetic
state of MnSi.  Rather I simply assume that long wavelength spin 
correlations persist over a wide range of pressure above $P_C$ and then predict their 
effect on current relaxation.           

I begin by reviewing the MnSi Fermi surface and then give
the qualitative explanation for the temperature-dependent
resistivity.  An explicit calculation of the resistivity, based on
a single-particle effective scattering rate, follows below and some discussion of temperature
 crossovers away from the $T^{3/2}$ regime is given near the end of the article.

Jeong and Pickett\cite{jeon04} (also see Ref. \onlinecite{tail86})  describe the Fermi surface of
MnSi as consisting of three sheets: one is a $\Gamma$-centred pocket
(that is not discussed here) and the other two are ``jungle gym''
open sheets with necks on the faces of the cubic Brillouin zone.
The latter two, which I label $\alpha$\ and $\beta$ as in Fig. 1,
intersect one another at the cube faces, thus sharing an
elliptical intersection with the boundary plane.  (The degeneracy
at points on the cube face is protected by the symmetry of the B20
crystal.\cite{jeon04})
  Interestingly, the normal of each sheet makes an angle
with the normal to the cube face that is not equal to $\pi/2$.  Translational
symmetry is nonetheless preserved since one sheet merges smoothly into the other
as the cubic face is crossed.  An illustrative cross-section\cite{note_pick} of the
$\alpha$ and $\beta$ sheets is shown on the left side of Fig. 1 and
a zoom on their intersection is shown in the upper right.  At their intersection, the
 Fermi velocities for the two sheets point
in different directions\cite{note_eta}, making an angle $2\eta$ in
Fig. 1. Given the intimate connection between the $\alpha$ and
$\beta$ sheets, it seems likely that strong interband spin
correlations at the $\alpha-\beta$ intersection will exist
whenever intraband correlations do.

\begin{figure}
\begin{center}
\includegraphics[scale=0.34]{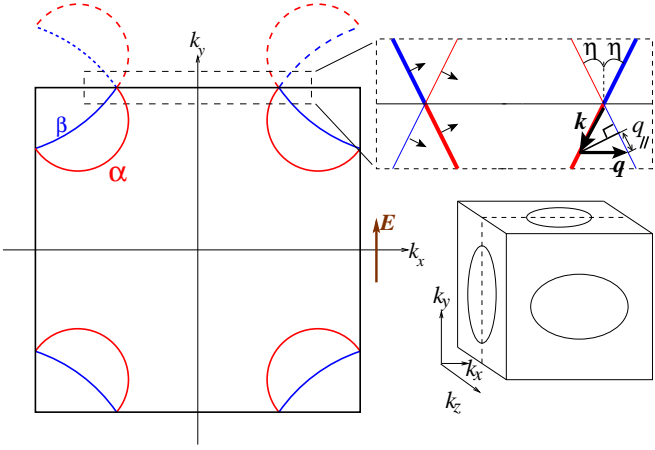}
\end{center}
\caption{\label{conepic} Colour online.  Left: 2D section of Brillouin zone
showing the $\alpha$ and $\beta$ jungle gym sheets of the MnSi
Fermi surface as inner and outer curves, respectively (red and blue, respectively, in colour version). 
The section is indicated by a dashed line on cube in the lower right. Upper right:  Zoom view of the
intersection of $\alpha$ and $\beta$ sheets.  Arrows on
left figure represent velocity while on the right a typical
small-$q$ interband scattering process is shown.  The thicker
lines of the F.S. have higher electron density in the applied
field ${\bf E}\parallel\hat{{\bf k}}_y$. Lower right: Surface of
Brillouin zone with elliptical $\alpha-\beta$
intersections indicated on cube faces.  Fig. 1 is meant
to illustrate qualitative features and is not
quantitatively accurate. Compare to Fig. 7 of Ref.
\onlinecite{jeon04}}
\end{figure}

If long-wavelength spin fluctuations that couple spins on the
$\alpha$ and $\beta$ sheets are important, then the ${\bf
k}$-space region shown in the zoom of Fig. 1 is of special
interest for transport. Here, an electron can be scattered by a
small-$q$ interband fluctuation from one sheet to the other. Since
the Fermi surface velocities make a finite angle at the
intersection, the current of the scattered electron changes
significantly when this happens. Thus, unlike scattering from
intraband small-$q$ spin fluctuations, the effect of interband
scattering on current transport does not vanish with $q^2$ due to
the $1-\cos\theta$ vertex factor. For similar reasons, the
Landau damping frequency of the spin fluctuations is independent
of $q$, rather than linear in $q$ as it is for intraband
fluctuations.

It becomes apparent that the phase-space for
 scattering by interband ferromagnetic spin-fluctuations in MnSi is similar
 to that for scattering from spin fluctuations in a nearly antiferromagnetic
 3D metal.  In both cases there is a line of points on the F.S.
 (the $\alpha-\beta$ intersections play the role of the hot spots)
near which electrons are susceptible to scattering by small-$q$
fluctuations in the order parameter. The damping rate of the spin
fluctuation is independent of $q$ and the scattering process
effectively relaxes the current when $q=0$.  It is therefore not
surprising that a $T^{3/2}$ term in the resistivity, which is
well-known for nearly
antiferromagnetic-metals\cite{hert76,mill93,mori95,kamb96} should
also occur in MnSi.

The simple preceding argument captures the main qualitative result
of this article. I now calculate the $T^{3/2}$ coefficient, taking
advantage of the analogy with the antiferromagnetic case.  To describe 
the long wavelength spin fluctuations I 
adopt the following model spin-susceptibility\cite{lonz85},  
 \begin{equation}
\chi({\bf q},\omega)=\frac{c}{-i\omega/\omega({\bf q})+\xi^{-2}+q^2}.
\label{chi}
\end{equation}
The dimensionless correlation length $\xi$ (distances are in units
of lattice constants and $\hbar=1$ unless otherwise stated) is
assumed to be much larger than unity and independent of $T$ at low $T$.  The Landau
damping frequency $\omega({\bf q})$ is discussed below and $c$ is
a constant.  In this article, the temperature range of interest is
that for which the factor $\omega({\bf q})(q^2+\xi^{-2})$ is much
smaller than a typical thermal frequency $\omega$ at small $q$ and
much larger than it for $q=1$.

Explicit Fermi sheet labels are now introduced (spin labels are
never used in this letter) in the calculation of the Landau
damping rate.  (The main interest is interband scattering near the
$\alpha-\beta$ intersecton; expressions are general enough to
include the intraband calculation but the latter is done roughly,
i.e. for an isotropic band.) From the bubble diagram for the
susceptibility with one electron Green's function having momentum
${\bf k}$ on Fermi sheet $\nu$ and the other momentum ${\bf k +
q}$ on Fermi sheet $\nu^{\prime}$, the Landau damping frequency is found to
satisfy
\begin{equation}
\omega_{\nu\nu^{\prime}}({\bf q})\propto\bigg{[}\int_{\nu}
\frac{dS_{\bf k}}{v_{\nu{\bf k}}} \delta[{\bf v}_{\nu^{\prime}{\bf
k+q}}\cdot({\bf k+q})-\epsilon_f]\bigg{]}^{-1}. \label{landau}
\end{equation}
where the integral is over the $\nu$th sheet of the Fermi surface,
${\bf v}_{\nu{\bf k}}$ is the Fermi velocity at ${\bf k}$ on the
$\nu$th sheet and $\epsilon_f$ is the Fermi energy.

For the intraband case $\nu=\nu^{\prime}$, Eq. \ref{landau} gives
$\omega_{\nu\nu}({\bf q})\propto q$ for small $q$, while for
interband fluctuations involving non-intersecting Fermi sheets the
integral is zero for $q$ smaller than the narrowest separation of
the Fermi sheets. But if distinct sheets $\nu$ and $\nu^{\prime}$
share a line of points ${\bf k}$ at which ${\bf v}_{\nu{\bf k}}\neq {\bf
v}_{\nu^{\prime}{\bf k}}$, as the $\alpha$ and $\beta$ sheets do
in MnSi, then the Landau damping rate is finite for $q=0$. For the
Fermi surface of Fig. 1, the $q=0$ value of the integral in Eq.
\ref{landau} is $6\pi k_f/(v_f^2\sin 2\eta)$ where $2\pi k_f$ is the
circumference of the elliptical $\alpha-\beta$ intersection and
$v_f$ is the magnitude of the Fermi velocity at this intersection, taken to be the 
same for both sheets.

The Landau damping rate is $q$ independent for long wavelength
interband spin fluctuations in MnSi.  In this regard they resemble
antiferromagnetic spin fluctuations for which small deviations
away from the ordering wavevector ${\bf Q}$ do not affect the
damping rate.

Having the spin fluctuation propagator from Eq. \ref{chi}, the
quasiparticle scattering rate may be calculated to lowest order
from the self energy diagram containing this propagator and a free
electron Green's function.  However, the current relaxation rate
cannot be set equal to the quasiparticle scattering rate for
nearly ferromagnetic systems
 because vertex corrections to low-$q$ processes are important.  Since I am to use
a relaxation-time approximation to the Boltzmann equation, some
vertex correction factor must be inserted into the momentum
integrand of the self-energy in order to account for the fact that
certain scattering processes do not alter the current. I use the
vertex correction factor appropriate for elastic scattering, which
is
\begin{equation}
\Lambda_{\nu\nu^{\prime}}({\bf q})=1-\frac{{\bf
v}_{\nu^{\prime}{\bf k}^{\prime}}\cdot{\hat{\bf u}}}{{\bf
v}_{\nu{\bf k}}\cdot\hat{{\bf u}}}. \label{vertex}
\end{equation}
where ${\bf k}^{\prime}={\bf k+q}$ and $\hat{{\bf u}}$ is a unit
vector in the direction of the applied field. Directions for which
${\bf v}_{\nu{\bf k}}\cdot\hat{{\bf u}}=0$ are of no concern since
they do not contribute to the current.

The presence of $\Lambda_{\nu\nu}({\bf q})$ results in an extra
factor of $q^2$ in the $q$ integral for intraband scattering. (The
term that is linear in $q$ in Eq. \ref{vertex} vanishes in the
angular integral over $\hat{{\bf q}}$.) A similar $q^2$ factor
occurs for scattering by acoustic phonons.

For interband scattering between the $\alpha$ and $\beta$ sheets
close to their intersection, the vertex correction
$\Lambda_{\alpha\beta}({\bf q})$ is zero for the four cube faces
that are parallel to $\hat{{\bf u}}$ since the velocity component
parallel to the field is equal for the two sheets on these faces.
On faces perpendicular to $\hat{{\bf u}}$ interband scattering
reverses the velocity along the field, as seen in Fig. 1, so
$\Lambda_{\alpha\beta}(0)=2$.

Since $\Lambda_{\alpha\beta}(0)$ is large, interband scattering on
the faces orthogonal to $\hat{{\bf u}}$ relaxes the current even
when $q=0$. Again this is similar to antiferromagnetic
fluctuations, which effect a large velocity change by scattering
an electron through ${\bf Q}$. It seems unlikely that a better
treatment of vertex corrections would change this qualitative
result. A contribution to the current by electrons on cube faces
perpendicular to $\hat{{\bf u}}$ is possible in MnSi because the
$\alpha$ and $\beta$-sheet normals do not lie in the plane of the
zone boundary.  Generic open Fermi surfaces have normals parallel
to the faces at the boundary, which makes scattering on faces
perpendicular to the field irrelevant to transport in most metals.

The momentum integral in the interband self-energy calculation may
be carried out using coordinates shown in the zoom of Fig. 1.  A
convergent Fermi surface integral of a function of $q$ is done as
$\int dS_{\bf q} f(q)=2\pi\int q_{\parallel}dq_{\parallel}
f\left(\sqrt{q_{\parallel}^2+k^2\sin^2 2\eta}\right)$. I define
$\xi^{-2}_{\alpha\beta}(k)=\xi^{-2}_{\alpha\beta}(0)+k^2\sin^2
2\eta$ where the interband correlation length
$\xi_{\alpha\beta}(0)$ is expected to be different from the
corresponding intraband quantity
$\xi_{\nu\nu}(k)=\xi_{\nu\nu}(0)$, but to share the general
features noted above.  Also, $\xi_{\alpha\beta}^{-2}(k)$ is much
larger than $k_BT/\omega_{\alpha\beta}(0)$ when $k$ is of order
unity.

The current relaxation rate $\tau^{-1}_{\nu\nu^{\prime}}({\bf
k},\omega)$ associated with the scattering of electrons from band
$\nu$ with momentum ${\bf k}$ and energy $\omega$ to any momentum
or energy on band $\nu^{\prime}$ is given at zero
temperature by
\begin{equation}
\tau^{-1}_{\nu\nu^{\prime}}(k
,\omega)=g_{\nu\nu^{\prime}}\int_0^{\omega} d\epsilon \epsilon
\int \frac{dq q}{v_f} \frac{\Lambda_{\nu\nu^{\prime}}(
q)\omega_{\nu\nu^{\prime}}(
q)}{\epsilon^2+\omega_{\nu\nu^{\prime}}^2(
q)(\xi_{\nu\nu^{\prime}}^{-2}(k)+q^2)^2}. \label{tau}
\end{equation}
where $g_{\nu\nu^{\prime}}$ is a constant energy coupling. The
angular integral has removed directional dependence so that, for
example, $\Lambda_{\nu\nu}(q)=1-\hat{{\bf v}}_{\nu{\bf
k}}\cdot\hat{{\bf v}}_{\nu{\bf k}^{\prime}}=q^2/2$.

The frequency dependence of Eq. \ref{tau} can be found by power
counting.  The intraband case has $\Lambda_{\nu\nu}(q)\propto q^2$
and $\omega_{\nu\nu}(q)\propto q$ while $\xi^{-2}_{\nu\nu}(k)$ is
a constant much smaller than unity. The second term in the
denominator of the $q$ integral goes as $q^6$ so powers of
$\omega$ are assigned according to $q\propto\omega^{1/3}$.  Thus
the $q$ integral goes as $\omega^{-1/3}$ and $\tau^{-1}_{\nu\nu}(k
,\omega)\propto\omega^{5/3}$. This is a well known result for
current relaxation near a ferromagnetic transition.

In the case of interband scattering both
$\Lambda_{\alpha\beta}(q)$ and $\omega_{\alpha\beta}(q)$ are
constant for small $q$ and the $q$ integral depends on $k$ through
$\xi_{\alpha\beta}(k)$.  At the $\alpha-\beta$ intersection, the
current relaxation rate is 
\begin{equation}
\tau^{-1}_{\alpha\beta}(0,\omega)=\frac{\pi g_{\alpha\beta} \omega}{2 v_f},
\end{equation}
while far from the $\alpha-\beta$ intersection it is approximately given by 
\begin{equation}
\tau_{\alpha\beta}^{-1}(k,\omega)=\frac{g_{\alpha\beta}\omega^2}{v_f k_BT_0}\;\;\;\;\;\;\; k\approx k_f 
\end{equation}
where $k_BT_0\approx k_f^2 \omega_{\alpha\beta}(0)$ (recall that the temperature range of 
interest lies well below $T_0$).  The situation is identical to that of nearly 
antiferromagnetic metals: the relaxation rate is linear in $\omega$ at hot spots and
quadratic in $\omega$ elsewhere.  More precisely, the relaxation rate is linear in $\omega$ 
when $\omega_{\alpha\beta}(k)\xi_{\alpha\beta}^{-2}(k)<<\omega$.  The basic assumption of this 
work is that, for MnSi, there exists a low-temperature regime at pressures well above $P_C$ where this 
condition is satisfied by thermal $\omega$ and $k$ sufficiently close to the $\alpha-\beta$ 
intersection. 

A relaxation-time approximation is used to calculate $\rho(T)$.
This is valid at low $T$, where the impurity scattering rate $\tau_{el}^{-1}$ exceeds the
inelastic scattering rate, in nearly antiferromagnetic
metals\cite{rosc00}. Since the calculation of the resistivity proceeds exactly as in the 
antiferromagnetic case (described in the Appendix of Ref. \onlinecite{rosc00}, for example),
I only sketch it briefly. 

The nonequilibrium part of the quasiparticle distribution $\delta
n_{\nu{\bf k}}$ in an applied field ${\bf E}=E_0\hat{{\bf u}}$ is
assumed to be
\begin{equation}
\delta n_{\nu{\bf k}}=eE_0\big{(}-\frac{d
n_0}{d\epsilon_{\nu}}\big{)}{\bf v}_{\nu{\bf k}}\cdot \hat{{\bf
u}}\frac{1}{\tau_{in}^{-1}({\bf
k},\epsilon_{\nu})+\tau^{-1}_{el}}. \label{neq}
\end{equation}
with the resulting current
\begin{equation}
{\bf j}=e\sum_{\nu,{\bf k}}\delta n_{\nu{\bf k}} {\bf v}_{\nu{\bf
k}}. \label{current}
\end{equation}
Band-labels have been dropped from the inelastic
scattering rate $\tau_{in}^{-1}({\bf k},\epsilon_{\nu})$ since
interband scattering is assumed and the energy argument identifies
the initial band.  The Fermi function is written as $n_0$.

To lowest order in $\tau_{in}^{-1}(k,\epsilon_{\nu})\tau_{el}$,
Eq. \ref{current} gives
\begin{equation}
\frac{\rho(T)-\rho(0)}{\rho(0)}=\rho(0)\sigma_1(T)
\end{equation}
where
\begin{equation}
\sigma_1(T)=\frac{2e^2\tau^2_{el}}{(2\pi)^3}\sum_\nu\int
d\epsilon_{\nu}\bigg{(}-\frac{dn_0}{d\epsilon_\nu}\bigg{)}\int_\nu
\frac{dS_{\bf k}}{v_{\nu,{\bf k}}}\tau_{in}^{-1}({\bf
k},\epsilon_{\nu}). \label{sig_in}
\end{equation}

I substitute Eq. \ref{tau} into Eq. \ref{sig_in} and note that the ${\bf k}$ integral 
converges within a strip of width $k_f\sqrt{T/T_0}<<k_f$ about the $\alpha-\beta$
intersection. The result, written in standard units,
is
\begin{equation}
\sigma_1(T)=\Gamma\frac{ne^2\tau_{el}}{m}\bigg{(}\frac{T}{T_0}\bigg{)}^{3/2}
\label{sig_lowin}
\end{equation}
where
\begin{equation}
\Gamma=\tan\eta \frac{g_{\alpha\beta}}{\hbar\tau^{-1}_{el}}\left(\frac{k_BT_0}{\epsilon_f}\right)
\end{equation}
to within a factor of order unity and $n/m=\epsilon_f k_f/(\hbar\pi)^2$, $\epsilon_f=\hbar v_f k_f$.  Similar
to the antiferromagnetic case, there is a crossover to Fermi liquid behaviour for $T<T_c$ where 
$T_c=k_f^{-2}\xi^{-2}_{\alpha\beta}(0)T_0<<T_0$.  This means that
$(T_0/T_c)=k_f^{2}\xi^{2}_{\alpha\beta}(0)$ is the largest temperature range over which $T^{3/2}$-resistivity 
can occur in this model.  The data\cite{doir03}, which show $T^{3/2}$ resistivity spanning two decades at 
$P=2P_C$, thus require that the interband spin correlation length at the $\alpha-\beta$ intersection is at 
least of order ten atomic spacings at this pressure.  Although it would be unusual, the existence of such long 
spin correlations well away from critical pressure is not inplausible in MnSi, especially given that 
Binz {\it et al}'s model\cite{binz06} of the partially ordered state above $P_C$ appears to have static ferromagnetic spin 
correlation lengths of the required order.       
 
The $k$-averaged inelastic relaxation rate in Eq. \ref{sig_in} emerges from the expansion in small 
$\tau_{in}^{-1}(k,\epsilon_{\nu})\tau_{el}$.  However, if one uses the crude procedure
of simply {\it replacing} $\tau_{in}^{-1}(k,\epsilon_{\nu})$ by its $k$-average in the vicinity 
of the $\alpha-\beta$ intersection, then the calculation may be done as above without assuming that
impurity scattering is dominant.  The result is $\rho(T)\propto T^{3/2}$ even when 
$\tau_{in}^{-1}(0,k_BT)>>\tau^{-1}_{el}$.  The general statement is that $T^{3/2}$-resistivity occurs when 
current relaxation is dominated by scattering from long-wavelength interband spin fluctuations near 
the $\alpha-\beta$ intersection.

For nearly antiferromagnetic metals, one expects $\rho(T)-\rho(0)$ to vary as $T^{3/2}$ only for temperatures 
low enough that $\rho(T)-\rho(0) << \rho(0)$ (this agrees with recent 
data on heavy fermion metals \cite{rosc00,pagl06}). This is because the $T^{3/2}$-dependence only 
occurs while the impurity scattering rate is sufficient to replenish the nonequilbrium density at hot spots.  At higher $T$, 
the nonequilibrium density is removed from hot spots by rapid inelastic scattering and the resistivity is determined by scattering 
that occurs elsewhere on the Fermi surface.  On the other hand, measurements of MnSi see $T^{3/2}$ behavior up to 
temperatures for which $\rho(T)>>\rho(0)$.  This raises the question of
how could scattering at $\alpha-\beta$ intersections remain important if 
impurity scattering is unable to maintain the nonequilibrium density at these intersections.   

A proper answer to this question would likely require a
Boltzmann-equation analysis as done in Ref.
\onlinecite{rosc00} for antiferromagnetic metals, but an obvious 
suggestion is the following: scattering by intraband fluctuations replenishes the
electron distribution near the $\alpha -\beta$ intersection while
having little effect on the current.  I elaborate on this before concluding. 
Interband scattering tends to equilibrate
the densities on the $\alpha$ and $\beta$ sheets at
their intersection. Intraband scattering counters this effect by
redistributing electrons along each sheet in an effort to
equilibrate the density at the intersection with that far from it.
This intraband redistribution process can be effective even when
current relaxation by intraband scattering is not.  (Consider that 
interband scattering tends to cause the electron density at the intersection 
to vary on the scale $\delta k\approx k_f(T/T_0)^{1/2}<<k_f$ along the Fermi surface throughout the 
$T^{3/2}$ regime.  Intraband 
scattering processes with momentum transfer $q<<k_f$ could prevent this tendency, at least up to temperatures 
approaching $T_0$, while having negligible effect 
on the current.)  It thus seems possible that intraband scattering could be responsible
for the robustness of the $T^{3/2}$ behavior in MnSi.  I leave a detailed study of the interplay 
between intraband and interband scattering near Fermi sheet intersections to future work.

In summary, I have presented a potential explanation for non-Fermi
liquid resistivity in MnSi that is based on the usual model of a
nearly ferromagnetic metal but takes into account unusual
features of the Fermi surface.  The work suggests that interband
spin correlations near the Fermi sheet intersections are important
in MnSi, and that their effect on other properties
should be considered.  I thank A. Boonchun and P. Pairor for useful discussions.


\begin{thebibliography}{99}
\bibitem{note_pick} Fig. 1 is based on the majority spin Fermi surface of Ref. \onlinecite{jeon04}. The minority-spin
surface, though qualitatively similar, appears to have a much
smaller $\eta$. I know of no band-structure studies of the
high-pressure paramagnetic state and simply assume that the
crucial $\alpha - \beta$ intersection feature found for the
ferromagnetic state also occurs at high pressure.
\end{thebibliography}
\end{document}